\documentclass[twocolumn,showpacs,preprintnumbers,prb,fleqn,floatfix]{revtex4}

\usepackage{graphicx}
\usepackage{dcolumn}
\usepackage{bm}

\begin{document}

\title{The quantum Hall ferromagnet at high filling factors:\\A
magnetic field induced Stoner transition}
\author{B. A. Piot and D. K. Maude}
\affiliation{ Grenoble High Magnetic Field Laboratory, Centre
National de la Recherche Scientifique, B.P. 166, F-38042 Grenoble
Cedex 9, France \\}
\author{M. Henini}
\affiliation{School of Physics and Astronomy, University of
Nottingham, Nottingham, NG7 2RD, U.K.}
\author{Z. R. Wasilewski}
\affiliation{Institute for Microstructural Sciences, National
Research Council, Ottawa, Canada, K1A 0R6.}
\author{K. J. Friedland, R. Hey and K. H. Ploog}
\affiliation{Paul Drude Institut f\"{u}r Festk%
\"{o}rperelektronik, Hausvogteiplatz 5-7, D-10117 Berlin, Germany}
\author{A. I. Toropov}
\affiliation{Institute of Semiconductor Physics, Prosp.
Lavrentyeva 13, 630090 Novosibirsk, Russia}
\author{R. Airey and G. Hill}
\affiliation{Department of Electronic and Electrical Engineering,
University of Sheffield, Sheffield S1 4DU, U.K.}

\date{\today }

\begin{abstract}

Spin splitting in the integer quantum Hall effect is investigated
for a series of Al$_{x}$Ga$_{1-x}$As/GaAs heterojunctions and
quantum wells. Magnetoresistance measurements are performed at mK
temperature to characterize the electronic density of states and
estimate the strength of many body interactions. A simple model
with no free parameters correctly predicts the magnetic field
required to observe spin splitting confirming that the appearance
of spin splitting is a result of a competition between the
disorder induced energy cost of flipping spins and the exchange
energy gain associated with the polarized state. In this model,
the single particle Zeeman energy plays no role, so that the
appearance of this quantum Hall ferromagnet in the highest
occupied Landau level can also be thought of as a magnetic field
induced Stoner transition.

\end{abstract}

 \maketitle

\section {INTRODUCTION}

The integer quantum Hall effect (QHE),\cite{Klitzing80} observed
in the magnetotransport properties of a two-dimensional electron
gas (2-DEG) can be understood within the framework of a single
electron picture. The magnetic field quantizes the orbital motion
of electrons, so that the density of states consists of discrete,
well defined Landau levels, giving rise to the even integer QHE.
When the Zeeman energy is included, the spin degeneracy of each
Landau level is lifted, giving rise to the odd integer QHE. While
the single particle picture works well for the even integer QHE,
this is not the case for the odd integer QHE, where it misses
most, if not all, of the essential physics. Following the
pioneering theoretical work of Ando and Uemura,\cite{Ando74} the
important contribution of exchange interactions to the spin gap is
experimentally well
established.\cite{Nicholas88,Usher91,Leadley98}

Spectacular manifestations of many body effects have been
extensively reported for low filling factors, notably at filling
factor $\nu=1$, where the ground state is an itinerant quantum
Hall ferromagnet,\cite{Jungwirth} with an excitation spectrum
which is either a spin wave\cite{Bychkov} or a spin texture
excitation (Skyrmion).\cite{Fertig} The appearance of a quantum
Hall ferromagnet, is analogous to the Stoner transition used to
describe ferromagnetism in metals.\cite{Stoner} In contrast, at
high filling factors, in the limit of low magnetic field, the
Shubnikov de Haas oscillations reveal that the spin up and down
levels are degenerate, indicating that exchange interactions are
not sufficiently strong to induce a ferromagnetic order within the
highest occupied Landau level. In this work we investigate the
transition between these two limiting behaviors. This transition
is commonly referred to as the appearance of spin-splitting. We
show that this phenomenon can be understood by adapting the Stoner
condition to take into account the modified 2D density of states
in a magnetic field.

In the most extensively investigated system, namely
Al$_{x}$Ga$_{1-x}$As/GaAs heterostructures, the single particle
Zeeman energy is rather weak ($\sim0.3$ K/T) owing to the small
effective electronic g-factor of GaAs ($g^{*}=-0.44$
),\cite{Weisbuch} whereas the experimental spin gaps evaluated,
using for example thermal activation, are the order of $\sim6$~K/T
for the case of fully developed spin polarization at odd filling
factor. This demonstrates the importance of the exchange
enhancement of the spin gap, which is as a result often written as
$g_{ex}^{*}\mu_{B}B$, where $\mu$$_{B}$ is the Bohr magneton and
$g_{ex}^{*}$ is an exchange-enhanced effective g-factor first
proposed in Ref.[\onlinecite{Janak}]. The exchange-enhanced spin
gap depends on the spin polarization of the 2-DEG, \cite
{Ando74,Nicholas88} and therefore, this description is actually
restricted to spin polarized odd filling factors (see for example
the theory of Aleiner and Glazman).\cite{Aleiner95} When
decreasing the magnetic field, experiments suggests that the
enhanced spin gap collapses, with odd integer filling factor
minima suddenly disappearing from the longitudinal resistivity
($\rho_{xx}(B)$) at low fields.

A theoretical explanation for the collapse of spin splitting has
been proposed by Fogler and Shklovskii,\cite {Fogler95} who
predicted a second order phase transition in which exchange
interactions are destroyed by disorder in the same way as
temperature destroys ferromagnetism in the mean field theory. More
precisely, when the spin gap attains the same order of magnitude
as the disorder-induced Landau level broadening, the spin
polarization at odd filling factor is reduced. This leads to the
destruction of the polarization (exchange) part of the enhanced
spin gap which can be qualitatively understood within the early
description of Ando and Uemura.\cite{Ando74} Experimentally, the
evolution of Fogler and Shklovskii's order parameter $\delta\nu$,
which is the filling factor difference between two consecutive
$\rho_{xx}$ maxima related to spin up and down Landau levels,
shows a universal behavior versus both temperature and electron
density after judicious re-scaling.\cite
{Wong97,Shickler97,Leadley98} Our experiments also confirms a
collapse of the exchange-enhanced spin gap with decreasing
magnetic field (Section \ref{SectionIVB}).

For a quantitative experimental description, parameters are needed
to accurately describe the competition between exchange
interactions and disorder. This is what is done in Ref.
[\onlinecite {Leadley98}] in which disorder is extracted from an
analysis of the low field Shubnikov de Haas oscillations, and
exchange is estimated from transport measurements of the spin gap
at higher fields, when spin splitting is fully resolved. From this
an empirical critical filling factor is extracted for the onset of
the collapse of spin splitting.

Here we propose to tackle the problem from an \emph{intuitively}
different point of view. Instead of starting from a spin polarized
situation in which the spin gap is open as when discussing the
collapse of spin splitting, we propose to tackle the phase
transition from the other side, by considering what happens to an
unpolarized (paramagnetic) 2-DEG, corresponding to the
$\delta\nu=0$ phase in the Fogler and Shklovskii approach, when
increasing magnetic field.

Spin-splitting can then be seen as a transition from a
paramagnetic state to a ferromagnetic state within a single Landau
level at odd filling factors. For this transition to occur we have
to compare the energy cost, due to the disorder induced Landau
level broadening, of populating higher energy levels by flipping
spins (inversely proportional to the density of states at Fermi
level, $D(E_{F})$) and the exchange energy gain associated with
the polarized state. At zero magnetic field this condition is
nothing other than the well-known Stoner condition for
ferromagnetism in metals.\cite{Stoner} In a magnetic field,
$D(E_{F})$ increases as $B/\Gamma $, where $\Gamma$ characterizes
the Landau level broadening. Increasing magnetic field increases
the density of states (Landau level degeneracy), thus lowering the
energy cost for flipping spins sufficiently to induce a
ferromagnetic state. For a quantitative comparison with
experiment, estimations of the Landau level broadening $\Gamma$
and the strength of the exchange energy are required.

The rest of this paper is organized as follows. In Section
\ref{SectionII} a simple model for the appearance of spin
splitting, based on the Stoner condition for ferromagnetism is
presented. The Landau level broadening is evaluated in Section
\ref{SectionIII} from an analysis of the low magnetic field
Shubnikov de Haas oscillations. The exchange energy is derived in
Section \ref{SectionIV} using the theoretical calculation of the
electronic spin susceptibility of Attaccalite \textit{et
al.}\cite{Attaccalite}. The calculated spin susceptibility and
exchange parameter are shown to be in good agreement with
experiment. Finally, in Section \ref{SectionV}, the Landau level
broadening and exchange energy are used to predict the critical
magnetic field for the appearance of spin splitting. A comparison
with the experimental data, shows good agreement for a number of
samples covering a wide range of densities and mobilities.

\section {MODEL FOR THE APPEARANCE OF SPIN SPLITTING}\label{SectionII}

\begin{figure}[tbp]
\includegraphics[width=0.9\linewidth,angle=0,clip]{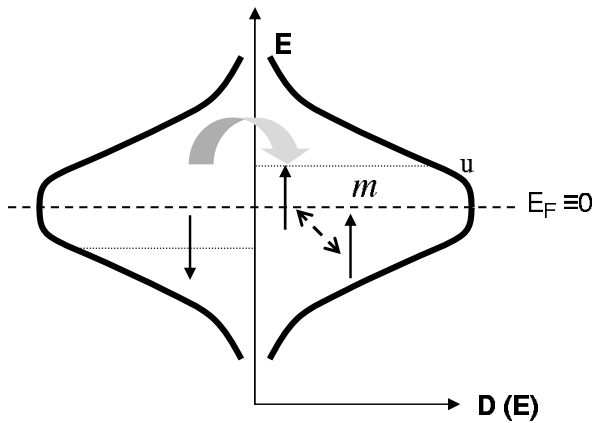}
\caption{\label{fig1} Schematic representation of the process for
spin splitting.}
\end{figure}

We propose here that exchange interactions drive a transition from
a paramagnetic state, corresponding to non spin-split Shubnikov de
Haas oscillations, to a ferromagnetic state, corresponding to spin
resolved Shubnikov de Haas. Strictly speaking, the transition is
driven by the density of states, via the Landau level degeneracy
$eB/h$, rather than the exchange energy, which is actually
magnetic field independent, to a good approximation, at high
filling factors (see Section \ref{SectionIVB}). The system is
described considering the highest occupied Landau level at odd
filling factor, so that the starting situation is when the system
is unpolarized and the Fermi level lies in the center of the
degenerate spin up and down sub levels of a given Landau level.
This is an approximation, since there can be a small residual spin
polarization due to the Zeeman splitting. However, in GaAs, at low
magnetic fields, the Zeeman energy is small compared to both
Landau level broadening and exchange energy, so it is a good
approximation to consider the limit of zero Zeeman energy. Within
this framework the appearance of a ferromagnetic state requires
that the energy cost of populating higher energy levels by
flipping spin, should be less than the gain in exchange energy,
which therefore stabilizes the newly polarized state. This
situation is depicted schematically in Fig.\ref{fig1}.

If we consider flipping spins to fill the spin up Landau level to
an energy $u$ above the initial Fermi energy, the energy cost can
be expressed as,
\begin{equation}\label{eq1}
\\E_{flip}= 2\int_{0}^{u}D(E) E~dE.
\end{equation}
To simplify, we have defined the zero of energy such that
$E_{F}=0$. It is possible to obtain a simple expression if we
assume that the density of states, $D(E)$, in this expression can
be approximated by its value at the Fermi level. Proceeding in
this way, we implicitly suppose that the Landau level is
rectangular shaped, which should be relevant only for small values
of $u$. If the temperature is zero and $m=2\int_{0}^{u}D(E)
dE\approx2D(E_{F})u$ is the spin polarization created by this
process within Landau level $N$, then Eq.(\ref{eq1}) can be
written as,
\begin{equation}\label{eq2}
\\E_{flip}=\frac{1}{4D(E_{F})}m^{2}.
\end{equation}
The factor 4 in Eq.(\ref{eq2}) simply arises from the definition
of $m$. This expression confirms the intuitive idea that the
larger the density of state is, the smaller the energy cost of
flipping spins will be, simply because more states are available
at lower energies. Calculations assuming a Gaussian form for the
Landau level show this approximation differs from the exact result
of Eq.(\ref{eq1}) by less than 5 \%, provided $m <0.425$.

At the same time this spin flip process leads to an increased
population of the spin up Landau level, giving rise to an exchange
energy gain which can be written,
\begin{equation}\label{eq3}
\\E_{X}=\frac{X_{N}}{4}m^{2}
\end{equation}
where $X_{N}$ is the exchange energy between 2 spins in Landau
level $N$. The quadratic dependence on $m$ results from the many
body nature of this interaction. Following this model in which we
only have two competing energies, it will be energetically
favorable for spins to flip when,
 \begin{equation}\label{eq4}
\\\frac{1}{D(E_{F})}=X_{N}.
\end{equation}
When the Fermi level lies the center of a Landau level, $D(E_{F})$
increases as $B/\Gamma $, due to the eB/h Landau level degeneracy
and the disorder-induced broadening $\Gamma$. The underlying idea
is that when increasing the magnetic field we reach a sufficient
density of states at the Fermi level to reduce the spin flip
energy cost, so that the spin system eventually becomes
``ferromagnetic''.

To check the validity of such a model, one must evaluate
quantitatively how the density of states varies with magnetic
field. For this a description of the shape and width of the Landau
levels is required.

\section {DENSITY OF STATES IN A MAGNETIC FIELD}\label{SectionIII}

\subsection {Low field Shubnikov de Haas analysis}\label{SectionIIIA}

\subsubsection {Formalism}

In a magnetic field, the diagonal conductivity $\sigma_{xx}$ of
the 2-DEG can be written,
\begin{equation}\label{eq5}
\\\sigma_{xx}=\frac{n_{s}e^{2}\tau_{tr}}{m^{*}(1+\omega_{c}^{2}\tau_{tr}^{2})}
\end{equation}
where $n_{s}$ is the electron density, $\tau_{tr}$ the transport
scattering time, $m^{*}$ the electron effective mass, and
$\omega_{c}$ the cyclotron frequency.

In the limit of a zero temperature and if the magnetic field is
large enough to satisfy $\omega_{c}\tau_{tr}\gg 1$ , this
simplifies to give,
\begin{equation}\label{eq6}
\\\sigma_{xx}\approx\frac{e^{2}D(E_{F})}{m^{*}\tau_{tr}\omega_{c}^{2}}.
\end{equation}
For high mobility samples, the Hall resistivity is usually much
larger than the longitudinal resistivity provided
$\omega_{c}\tau_{tr}\gg 1$, so that the 2-dimensional resistivity
tensor gives $\rho_{xx}\approx\sigma_{xx}\rho_{xy}^{2}$. In the
non spin-split low magnetic field region, $\rho_{xy}$ is
approximatively linear in B, and from Eq.(\ref{eq6}) the field
dependance of $\rho_{xx}$ can be written,
\begin{equation}\label{eq7}
\\\rho_{xx}\propto\frac{1}{\tau_{tr}}D(E_{F}).
\end{equation}
At low magnetic field, the variation $\Delta\rho$ of the
longitudinal resistivity around the zero field resistivity
($\rho_{0}$) can be calculated by expanding the quantized density
of states as a Fourier series. The amplitude of the fundamental
oscillatory term gives, in the zero temperature limit, the
following expression for $\Delta\rho / \rho_{0}$,
\begin{equation}\label{eq8}
\\
\frac{\Delta\rho}{\rho_{0}}=Ae^{{\frac{-\pi}{\omega_{c}\tau_{q}}}}.
\end{equation}
This expression, which is a good approximation provided that
$\Delta\rho/\rho_{0}<1$, is obtained for the particular case of
Lorentzian Landau level and is nothing more than the disorder
damping term in the Lifshitz Kosevitch formula.\cite{LK} Here
$\tau_{q}$ is the quantum life time related to the width of the
Lorentzian Landau level by $\tau_{q}=\hbar/2\Gamma_{L}$ . $A$ is a
constant whose value depends on the power relation between the
resistivity and the density of state (see below).

The same calculation for a Gaussian Landau level (defined here
with a full width at half maximum of $2\sqrt{\ln(2)\Gamma_{G}}$ )
gives,
\begin{equation}\label{eq9}
\\\frac{\Delta\rho}{\rho_{0}}=Ae^{{\frac{-\pi^{2}\Gamma_{G}^{2}}{\hbar^{2}\omega_{c}^{2}}}}
\end{equation}
The expression is similar  to that for a Lorentzian Landau level
except that the B dependence in the exponential function is in
$1/B^{2}$ if the Landau level width is field independent.

Plotting the experimental $\Delta\rho/\rho_{0}$ on a logarithmic
scale versus $1/B$ or $1/B^{2}$ can give information concerning
the shape of the Landau level. A linear behavior of
$\ln(\Delta\rho/\rho_{0}$) when plotted versus $1/B$  (so-called
Dingle plot) indicates Lorentzian Landau levels \cite{Gaussian},
whereas a linear behavior when plotted versus $1/B^{2}$ suggests
Gaussian Landau levels. In both cases the slope of these curves
gives the (field-independent) width of these levels. This
theoretical treatment has however to be confirmed by checking the
value of the intercept at $1/B=0$, noted here as $A$. An intercept
of $A=2$ is theoretically expected when the resistivity is
directly proportional to $D(E_{F})$. If however, the $\rho_{xx}$
dependence on $D(E_{F})$ is quadratic (\textit{i.e.}
$1/\tau_{tr}\propto D(E_{F})$ in Eq.(\ref{eq7})), the Fourier
expansion exhibits an additional factor 2, which leads to an
intercept of $A=4$. These issues are discussed in more detail in
the seminal paper by Coleridge.\cite{Coleridge96}

\subsubsection {Low temperature magnetotransport}\label{SectionIIIA2}

\begin{figure}[tbp] 
\includegraphics[width=0.9\linewidth,angle=0,clip]{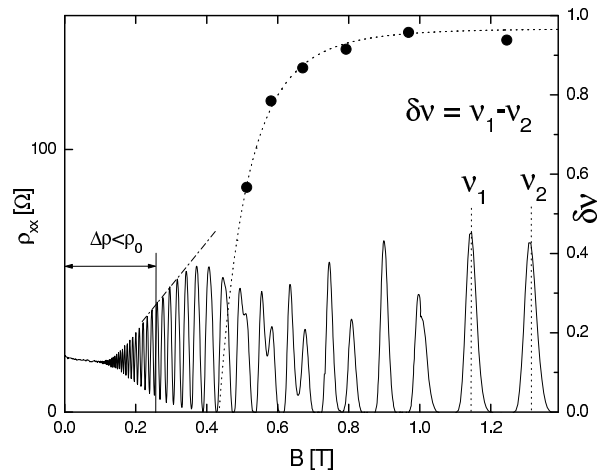}
\caption{\label{fig2} Magneto resistance trace for sample NU
1783a. $\delta\nu$ (full circles) as a function of magnetic field.
Dotted line is the extrapolation of $\delta\nu$ used to determine
$B_{ss}$ as explained in Section \ref{SectionV}. Vertical solid
line delimits the magnetic field region for which
$\Delta\rho/\rho_{0}<1$. Dash-dotted straight line emphasizes the
linear behavior of $\rho_{xx}^{peak}(B)$ at higher magnetic
fields.}
\end{figure}

\begin{figure}[tbp] 
\includegraphics[width=0.7\linewidth,angle=0,clip]{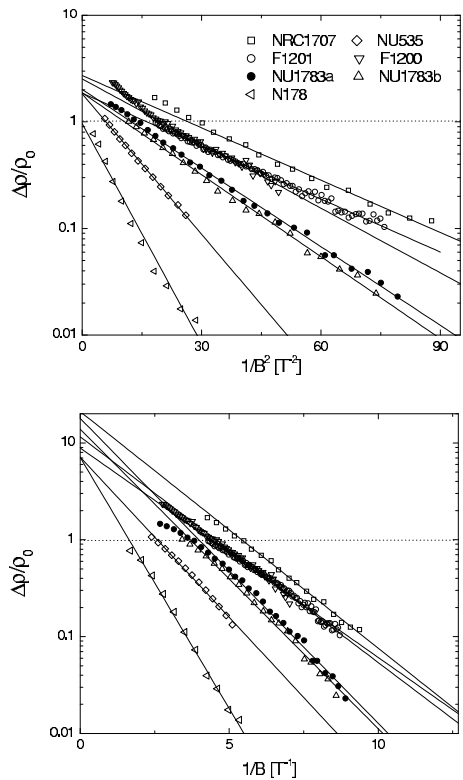}
\caption{\label{fig3} Dingle plots. (a) $\Delta\rho/\rho_{0}$
versus $1/B^{2}$ and (b) $\Delta\rho/\rho_{0}$ versus $1/B$. Note
log scale for vertical axes. Straight lines are linear fits to the
data points in the region $\Delta\rho/\rho_{0}<1$ delimited by the
dotted horizontal dotted line.}
\end{figure}

We have performed mK temperature magnetotransport measurements on
different hetero junctions(HJ) and quantum wells (QW), using a
standard low frequency lock-in technique. Representative
longitudinal resistivity ($\rho_{xx}$) versus magnetic field data
obtained for one of the samples is shown in Fig.\ref{fig2}. The
variation of the longitudinal resistivity around the zero field
value ($\Delta\rho/\rho_{0}$) has been directly extracted from the
low field Shubnikov de Haas data at 50mK. ($\Delta\rho/\rho_{0}$)
is plotted on a logarithmic scale versus ($1/B^2$) in
Fig.\ref{fig3}(a), and, for comparison, versus ($1/B$) in
Fig.\ref{fig3}(b), for all the samples investigated.

At $T=50$ mK, the smearing imposed by the Fermi-Dirac statistics
is small enough to ensure that $\Delta\rho$ can be analyzed using
a zero temperature formalism. In the samples studied, the Dingle
plots are found to decrease as $1/B^{2}$ rather than as $1/B$(see
Fig.\ref{fig3}). While the linear fits are reasonable in both
cases, they are slightly better in the $1/B^{2}$ plots
(Fig.\ref{fig3}(a)). The data in the $1/B$ plots has a noticeable
curvature (Fig.\ref{fig3}(b)). In addition, the intercepts in the
$1/B^{2}$ plots ($1<A<2.7$) are close to one of the theoretically
expected values, $A=2$. This is not the case for the $1/B$ plots,
with intercepts $7<A<21$, far from the theoretically expected
value of either $A=2$ or $A=4$.

This suggests, that for our samples, the density of states is
better described using a Gaussian form with field-independent
width for the Landau levels. Within this model, the intercept
$A\simeq2$, implies that $\rho_{xx}$ is directly proportional to
$D(E_{F})$. We have also checked that an identical result is
obtained when filtering the raw data with a band pass centered at
the fundamental Shubnikov de Haas frequency. The $1/B^{2}$
decrease (a Gaussian form for the Landau levels) therefore
provides a better description of our data. From Eq.(\ref{eq7})
this result implies, as proposed in Ref.[\onlinecite{Coleridge94}]
that the transport scattering time does not depend on $D(E_{F})$.

A complementary study to the low field Shubnikov de Haas analysis
can be performed by analyzing the values of the resistivity peaks,
$\rho_{xx}^{peak}(B)$, in the regime
$\Delta\rho/\rho_{0}>1$.\cite{Coleridge94,Stormer} In this regime,
we observe a linear increase of $\rho_{xx}^{peak}(B)$ (visible in
the $\rho_{xx}$ trace for sample NU1783a in Fig.\ref{fig2}). When
$\Delta\rho/\rho_{0}>1$, Eq.(\ref{eq8}) and Eq.(\ref{eq9})  are no
longer valid, because the deviation of the density of states from
the zero field value becomes too large to be described by the
fundamental oscillatory term in the Fourier series. However,
Eq.(\ref{eq7}) is still valid and thus the field dependence of the
resistivity peaks can provide information on
$\frac{1}{\tau_{tr}}D(E_{F})$. The less stringent condition to
account for the observed linear dependence of
$\rho_{xx}^{peak}(B)$, is as proposed in
Ref.[\onlinecite{Coleridge94}], that the inverse of the transport
relaxation time and the Landau level broadening have the same
magnetic field dependence. For the sample studied in
Ref.[\onlinecite{Coleridge94}], this dependance was in square root
of B, while here the low field Shubnikov de Haas (Dingle) analysis
suggests both parameters are field-independent. The latter case is
fully compatible with the observed linear dependence of
$\rho_{xx}^{peak}(B)$.

The theory of Raik and Shabazyan\cite{Raik} on the Landau level
shape in the presence of long range scattering, suggests Gaussian
Landau levels with field independent width in the regime where
$a_{corr}\gg R_{cycl}$ (where $a_{corr}$ is the correlation length
of the disorder potential, and $R_{cycl}$ the cyclotron radius)
and with a square root of B dependence when $ R_{cycl}\gg a_{corr}
$. Our results are consistent with a field-independent broadening
of the Gaussian Landau levels suggesting that $a_{corr}\gg
R_{cycl}$ in our samples. While this is an interesting point
discussed in Refs.[\onlinecite{Coleridge96,Raik}], it is not the
direct focus of the present work.

\subsection {Estimation of the Landau level width}

To evaluate the $D(E_{F})$ appearing in Eq.(\ref{eq4}), one needs
an estimation of the total Landau level width $\Gamma$, which
include all states, both delocalized and localized. Estimating
$\Gamma$ from the low field Shubnikov de Haas oscillations is
\textit{a priori} not possible because in this region delocalized
states from different Landau levels overlap, preventing the
resistance from going to zero. However, at slightly higher fields,
in the region just before spin-splitting appears, the proximity to
the first zero resistance states at even filling factors can be
used to estimate $\Gamma$. The width of the $\rho_{xx}$ minima
(zero resistance states) depends on the ratio
$\Gamma_{dl}/\Gamma$, where $\Gamma_{dl}$ characterizes the width
of the extended (delocalized) states.

\begin{figure}[tbp] 
\includegraphics[width=0.9\linewidth,angle=0,clip]{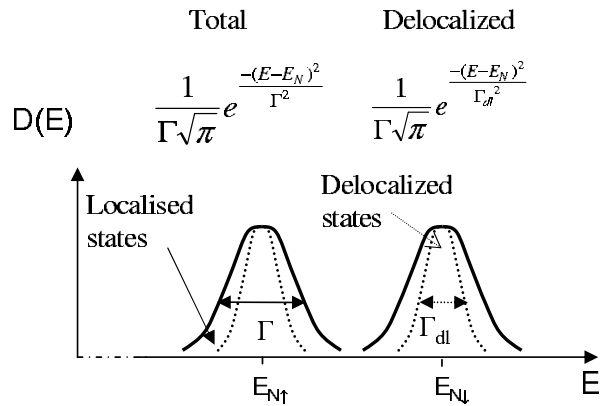}
\caption{\label{fig4} Schematic representation of the density of
states in a magnetic field used in a simple model to estimate
$\Gamma$ in Section \ref{SectionIII}.}
\end{figure}

Starting from the formalism presented in section
\ref{SectionIIIA}, we construct a simple model in the zero
temperature limit, assuming that the density of states can be
modeled as a set of Gaussian functions of width $\Gamma$, with
sub-Gaussians delimiting delocalized states near the center of the
Landau level, of width $\Gamma_{dl}$. This is schematically
depicted in Fig.\ref{fig4}. $\Gamma_{dl}\equiv\Gamma_{G}$, is
determined from the slope of the $\ln(\Delta\rho/\rho_{0})$ versus
$1/B^{2}$ plots as described in section \ref{SectionIIIA}. The
values of $\Gamma_{dl}$ for all the samples investigated are
summarized in table \ref{table1}. $\rho_{xx}$ can then be
calculated using $\rho_{xx} \propto D(E_{F})$. Knowing
$\Gamma_{dl}$, it is possible to determine $\Gamma$, for each
sample, by fitting to the $\rho_{xx}(B)$ data in the magnetic
field region just before spin-splitting appears. Proceeding in
this way we have estimated $\Gamma$ for the different samples
studied (see table \ref{table1}). Experimentally a Gaussian form
for the density of delocalized states gives by far the best fit to
our $\rho_{xx}(B)$ data. We have also tried to fit assuming the
region of delocalized states is delimited with a sharp cut-off
(mobility edge), as often proposed in the literature. While this
model is unable to correctly reproduce the shape of $\rho_{xx}(B)$
we can nevertheless roughly estimate $\Gamma$. The results are
comparable to those obtained with our Gaussian model, suggesting
that the Landau level width $\Gamma$ extracted here is to some
extent independent of the exact form used for the density of
delocalized states.

\begin{table*}[tbp]
 \begin{tabular}{cccccccc}
 Sample & Structure & $n_{s}$  & $\mu$ & $\Gamma_{dl}$~(K)  & $\Gamma$~(K) & $B_{ss}$~(T) & $g_{ex}^{*}$ \\
& & (10$^{11}$cm$^{-2}$) & (10$^6$cm$^{2}$/Vs) & & & \\
\colrule
F1201 & QW & 8.83 & 1.0 & $1.23\pm0.05$ & $2.3\pm0.1$ & $0.93\pm0.07$ & $5.3\pm1.5$ \\
F1200 & QW & 7.55 & 1.0 & $1.35\pm0.05$ & $1.9\pm0.1$ & $0.72\pm0.10$ & $5.7\pm0.4$ \\
NRC1707 & HJ & 1.64 & 5.8 & $1.10\pm0.05$ & $1.5\pm0.1$ & $0.26\pm0.05$ & $13.1\pm1.0$ \\
NU 1783a & HJ & 2.10 & 1.5 & $1.48\pm0.05$ & $2.2\pm0.1$ & $0.43\pm0.06$ & $13.0\pm1.0$ \\
NU 535 & QW & 2.13 & 0.11 & $2.00\pm0.05$ & $ 3.1\pm0.2$ & $0.74\pm0.12$ & $9.5\pm1.0$ \\
NU 1783b & HJ & 1.77 & 1.8 & $1.53\pm0.05$ & $ 2.3\pm0.2$ & $0.32\pm0.11$ & $13.7\pm0.7$ \\
N 178 & HJ & 1.32 & 0.32 & $2.50\pm0.05$ & $ -$ & $0.54\pm0.2$ & $16.0\pm1.5$ \\
 \end{tabular}
 \caption{Samples parameters: structure, electron
density ($n_{s}$), mobility at 50mK ($\mu$), width of delocalized
states ($\Gamma_{dl}$), total Landau level width ($\Gamma$),
magnetic field for the appearance of spin-splitting ($B_{ss}$)
discussed in Section \ref{SectionV} and exchange-enhanced g-factor
($g_{ex}^{*}$) discussed in Section \ref{SectionIVB}. For N 178 a
reliable estimation of $\Gamma$ was not possible.}\label{table1}
 \end{table*}

\section {ESTIMATION OF THE EXCHANGE INTERACTION}\label{SectionIV}

Analyzing the spin split Shubnikov de Haas oscillations we can
extract parameters which characterize exchange, such as, for
example, the enhanced-effective g-factors.\cite{Leadley98}
However, in this case the spin gap already exists, and, the system
is therefore already polarized. To describe an unpolarized 2-DEG
in a low field situation, it is better to estimate exchange via
the spin susceptibility of the electronic system, defined as the
response to an external magnetic field in terms of spin
magnetization. This is a good probe of exchange interactions
between electrons since exchange favors spin alignment and thus
enhances the spin susceptibility. We will see in section IV.B.
that these two different approaches to estimate the exchange
(enhanced-effective g-factors and spin susceptibility) are fully
consistent.

\subsection {spin susceptibility}\label{SectionIVA}

\subsubsection{Theoretical approach}

At zero magnetic field, the nature of the 2-DEG ground state is
strongly dependent on the electron density, and thus often
described as a function of the so-called density parameter
$r_{s}$, corresponding to the ratio of the Coulomb energy to the
kinetic energy. In 2D, $r_{s}$ scales as $1/\sqrt{n_{s}}$. At high
$r_{s}$, in the low density regime, the ground state remains a
topic of controversy, with a possible manifestation of a metal to
insulator transition (MIT) observed around
$r_{s}\sim10$.\cite{Kravchenko01} At low $r_{s}$ however, and
especially for the density range investigated here ($r_{s}<2$),
the density is sufficiently high to ensure that the zero field
ground state is a paramagnetic liquid. In this case, the
paramagnetic relative spin susceptibility $\chi/\chi_{0}$ of the
2-DEG can be derived from calculations of the $B=0$ ground state
energy as a function of density and polarization, for example
using quantum Monte Carlo simulations (see e.g. Tanatar and
Ceperley ).\cite{Tanatar89} Here, $\chi/\chi_{0}$ corresponds to
the ratio of the ``real'' spin susceptibility of interacting
electrons, $\chi$, to the non-interacting Pauli value $\chi_{0} $.
Similar calculations have been performed more recently by
Attaccalite \textit{et al.} \cite{Attaccalite}. Different
predictions arise from the different ways of estimating the total
energy of the ground state, and in particular the correlation
term. At low density, $\chi/\chi_{0}$ turns out to be a very
delicate quantity to estimate, and the various predictions are
significantly different.

\begin{figure}[b]
\includegraphics[width=0.9\linewidth,angle=0,clip]{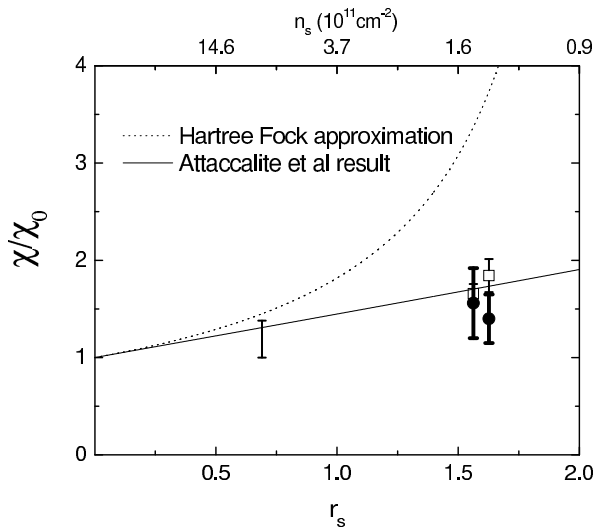}
\caption{\label{fig5} Theoretical spin susceptibility as a
function of the density parameter $r_{s}$. Hartree-Fock
approximation (dotted curve), Attaccalite \textit{et al.}
prediction (solid curve). Open squares are experimental values of
$\chi/\chi_{0}$ at $i=1$ determined by tilted-field experiments on
different samples. For the high density sample ($r_{s} \approx
0.7$) only an upper limit could be extracted. Full circles are
extrapolated ($B\rightarrow 0$) values of the paramagnetic
$\chi/\chi_{0}$.}
\end{figure}

Fortunately, for our rather high density systems ($r_{s}<2$), the
behavior of the predicted susceptibilities is more robust. Indeed,
predictions for small values of $r_{s}$ do not show large
discrepancies because in the limit $r_{s}\longrightarrow0$ they
all reach the ``basic'' Hartree-Fock approximation for spin
susceptibility, determined using the well- known Hartree-Fock 2D
ground state energy ($E_{g}=1/r_{s}^{2}-1.2004/r_{s}$, see e.g.
Ref.[\onlinecite{Isihara}]). This Hartree-Fock unscreened
susceptibility increases with increasing $r_{s}$, reflecting that
the role of exchange interactions becomes more important at low
density. This trend is common to a large majorities of predictions
over a very large $r_{s}$ domain. However, the Hartree-Fock
susceptibility diverges for $r_{s}\sim2.2$ because it does not
include the effects of the correlation energy and/or screening,
which prevents $\chi/\chi_{0}$ from diverging. Taking into account
these corrections, as done for instance in
Ref.[\onlinecite{Attaccalite}], the $r_{s}$ increase is still
present but much smoother. These different theoretical predictions
are plotted for comparison in Fig.\ref{fig5}.

\subsubsection{Experimental approach}

\begin{figure}[tbp]
\includegraphics[width=0.9\linewidth,angle=0,clip]{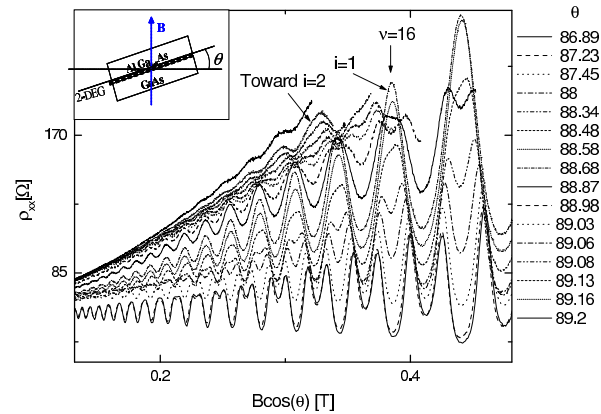}
\caption{\label{fig6} Magnetoresistance traces for sample NU 1783b
at different tilting angles. Angles are determined from the slope
of the low-field Hall resistance.}
\end{figure}

There are several experimental techniques to measure the electron
spin susceptibility of a 2-DEG (for a review see
Ref.[\onlinecite{Pudalov}]). For the particular case of a 2-DEG in
GaAs, significant results have recently been obtained with the
tilted field method by Zhu \textit{et al.}\cite{Zhu03}. The
principle of this method is to induce Landau level coincidences by
tilting the sample in a magnetic field. When tilting the sample,
the magnetic field perpendicular to the 2-DEG is reduced (see
inset of Fig. \ref{fig6}). We recall that the Zeeman gap depends
on the total magnetic field, whereas the cyclotron gap, which is
an orbital effect, depends only on the field component
perpendicular to the 2-DEG.

Therefore, by tilting the sample, for a given filling factor, we
can increase the relative weight of the Zeeman gap until the point
where the spin gap is equal to the cyclotron gap. This is what we
refer to here as the coincidence condition (note in the literature
coincidence condition often refers to the case where the spin gap
is half the cyclotron energy). Knowing the angle $\theta$ (defined
in the see inset of Fig. \ref{fig6}) which corresponds to this
situation, the product $ m^{*} g^{*}_{tilt}$ satisfying $ m^{*}
g^{*}_{tilt}=i2m_{e}\cos(\theta)$, where $i=1$ can be extracted.
If normalized using the band structure parameters
($m^{*}=0.068m_{e}$, $m_{e}$ being the electron rest mass, and
$|g^{*}|=0.44$ ), this $ m^{*} g^{*}_{tilt}$ is directly equal to
the relative spin susceptibility $\chi/\chi_{0}$ at $i=1$. The
parameter $g^{*}_{tilt}$ describe the total spin gap assuming it
can be written $g^{*}_{tilt}\mu_{B}B$ where B is the total
magnetic field consistent with the definition of the spin
susceptibility. The enhanced g-factor can also be defined using
the perpendicular component of the magnetic field, related to the
2-dimensional (orbital) nature of exchange.\cite{Leadley98}

Proceeding with a tunable density gated sample, \textit{Zhu et
al.}\cite{Zhu03} give an explicit density/polarization dependence
of the relative susceptibility for a range of density going from
$2\times10^{9}$ to $4\times10^{10}cm^{-2}$. The polarization
dependence is obtained from the observed i-dependence of
$\chi/\chi_{0}$, because different values of i correspond to a
different polarization of the system. The limit of zero
polarization can provide an estimation of the paramagnetic
relative susceptibility of the 2-DEG. Their results show the
experimental $\chi/\chi_{0}$ is much smaller than the Hartree-Fock
prediction and also that the $r_{s}$ trend is in good agreement
with the Attaccalite \textit{et al.} calculation, although the
experimental values are slightly smaller than predicted. Other
experimental data on different systems, mainly 2-DEG in Si, can
complete this comparison if they are rescaled in terms of $r_{s}$
(see \textit{e.g.} Ref.[\onlinecite{Pudalov}]), and show that this
trend is also respected in the lower $r_{s}$ region.

To check if Attaccalite \textit{et al.} prediction is
experimentally valid for GaAs in the region of interest here
($r_{s}<2$), we have performed tilted-field measurements on our
samples which cover a higher density range than the one in
Ref.[\onlinecite{Zhu03}]. Going towards high density should reduce
spin susceptibility and thus requires even larger tilt angles, and
thus higher magnetic fields. Transport measurements were performed
using standard low frequency lock-in technique under magnetic
fields up to 23~T, in a dilution fridge (base $T\approx50$~mK)
equipped with an ``\textit{in-situ}'' rotating sample holder.

In Fig.\ref{fig6} we plot $\rho_{xx}(B)$, measured at $T=50$~mK,
and at various tilt angles, for sample NU1783b. Note different
traces are ``naturally'' shifted upwards, due to the effect of the
increasing in plane magnetic field which leads to a positive
magnetoresistance.\cite{Dolgopolov} The angles were accurately
determined from the slope of the low magnetic field Hall
resistance. As expected, we find that coincidence is reached for
even larger tilt angles than in the \textit{Zhu et al.}
experiment.\cite{Zhu03} This is consistent with previous
measurements in this higher density range.\cite{Leadley98} For
example in sample NU1783b coincidence is reached at approximately
$88.7^{\circ}$ (see Fig.\ref{fig6}). The fact that we can use
tilted field to tune the Zeeman energy is not inconsistent with
our simple model in which the single particle Zeeman energy is
assumed to be negligible in the $\theta=0^{\circ}$ configuration.
The large tilt angle required, implies that an in plane magnetic
field approximately two order of magnitude larger than the
perpendicular magnetic field component is necessary to achieve
Landau level coincidence.

If we consider the situation at even filling factor, for example,
at $\nu=16$ indicated by the arrow in Fig.\ref{fig6}, we see that
when increasing $\theta$ the gap at even filling factor initially
closes, until the spin up and down sub-levels of adjacent Landau
levels coincide, here at about 88.7 degrees. Tilting further, we
are able to make the levels cross, confirming we do indeed obtain
a coincidence of Landau levels. We can see that the gap re-opens
at even filling factors, and then closes at odd filling factors
probably reaching a second coincidence in which the spin gap is
twice the cyclotron gap. The coincidence condition depends only
weakly on the perpendicular magnetic field. In other words, a
single tilt angle $\theta$, provides coincidence nearly
simultaneously at all observed even filling factors. This means
that the polarization dependance of $\chi/\chi_{0}$ is weak, more
precisely sufficiently weak, that its effects are suppressed by
Landau level broadening.

Determining the coincidence condition from the $\rho_{xx}$ trace,
the corresponding products $ m^{*} g^{*}_{tilt}$ have been
extracted for two samples measured (1783b and N178). For the high
density samples, coincidence occurs at higher tilt angles
($\theta>88.81^{\circ}$), and, due to sample mobility and magnetic
field limitations, coincidence cannot be reached. Nevertheless, we
can extract an upper limit for the product $ m^{*} g^{*}_{tilt}$
of sample 1200, showing that the susceptibility is indeed slightly
lower for this higher density sample. The values of $ m^{*}
g^{*}_{tilt}$, normalized by the product of the band structure
values $ m^{*}=0.068m_e$ and $|g^{*}|=0.44$, are plotted versus
$r_{s}$ in Fig.\ref{fig5} (open squares).

The paramagnetic spin susceptibility $\chi/\chi_{0}$ can then be
estimated from the i=0.5, 1 (and 1.5 when it is observed)
conditions at a given perpendicular magnetic field, assuming the
i-dependance of spin susceptibility is linear as observed in
Ref.[\onlinecite{Zhu03}]. Such a linear extrapolation induces
large error bars, because in the best situation we only have 3
values for $\chi/\chi_{0}(i)$, themselves subjected to error. As
the tilt angle required for different coincidence conditions is
roughly the same for all filling factors, extrapolation at
different perpendicular magnetic field gives similar results. The
differences are within the error bars induced by the linear
extrapolation. While, the value of $\chi/\chi_{0}$ at i=1 is
subject to less uncertainty than the extrapolated paramagnetic
$\chi/\chi_{0}$, rigourously it is the latter which is required
for a comparison with theory.

The paramagnetic $\chi/\chi_{0}$ are plotted versus $r_{s}$ for
sample 1783b and N178 in Fig.\ref{fig5} (closed circles). They are
consistent with the Attaccalite \textit{et al.} prediction,
smaller at most by 20 \%. The experimental values in
Ref.[\onlinecite{Zhu03}] are also smaller than the theoretical
prediction. However, the calculation of Attaccalite \textit{et
al.} is performed for an ideal 2-DEG, whereas experimentally the
finite thickness of a real 2-DEG is expected to reduce spin
susceptibility. Very recent finite thickness corrections to the
Attaccalite \textit{et al.} calculation have been performed by De
Palo \textit{et al.}\cite{Depalo}, showing these effects indeed
lead to a reduction of the theoretical values, reconciling them
with experiments (see also Ref.[\onlinecite{Zhang}]). Furthermore,
spin susceptibility measurements on system similar to the one used
in Ref.[\onlinecite{Zhu03}] have been recently extended to the
high density regime down to $r_{s}=0.8$,\cite{Tan} showing good
quantitative agreements with the finite thickness correction in
this density regime.\cite{Depalo}

What is important for this work, is that the Attaccalite
\textit{et al.} calculation provide a relevant estimation of spin
susceptibility in our samples, and therefore we can use this
theory to derive our exchange parameter.

\subsection {Exchange parameter versus density}\label{SectionIVB}

We now have to draw a formal link between the relative spin
susceptibility and the exchange parameter $X_{N}$ defined in
Section \ref{SectionII}. As $\chi/\chi_{0}$ is calculated for the
case of zero magnetic field, our condition for the appearance of
spin splitting should be applied in the limit $B=0$, which gives
$X_{\infty}D(E_{F})_{B=0}=1$.

As this condition correspond to a zero-field ``ferromagnetic''
state, we can define,
\begin{equation}\label{eq10}
\\ \frac{\chi}{\chi_{0}}=\frac{1}{1-X_{\infty}D(E_{F})_{B=0}}.
\end{equation}
Here $D(E_{F})_{B=0}$ is the zero-field density of states of the
2-DEG, for one spin orientation, i.e. $m^{*}/2\pi\hbar^{2}$. Here
$m^{*}$ is taken to be equal to the accepted GaAs band edge value
of $0.068m_{e}$. The zero field exchange parameter $X_{\infty}$ is
then obtained from the Attaccalite \textit{et al.} spin
susceptibility $(\chi/\chi_{0})_{Att}$ as a function of density,
\begin{equation}\label{eq11}
\\ X_{\infty}(n_{s})
=\frac{(\frac{\chi}{\chi_{0}})_{Att}(n_{s})-1}{(\frac{\chi}{\chi_{0}})_{Att}(n_{s})D(E_{F})_{B=0}}.
\end{equation}

\begin{figure}[tbp]
\includegraphics[width=0.9\linewidth,angle=0,clip]{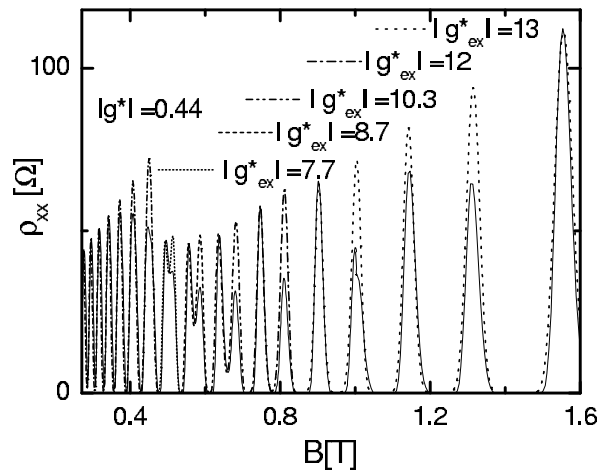}
\caption{\label{fig7} Measured $\rho_{xx}$ versus magnetic field
for sample NU1783a at $T=50$~mK (solid line). Calculated
$\rho_{xx}$ taking into account exchange interactions with a
phenomenological spin gap $g_{ex}^{*}\mu_{B}B$ (broken lines).}
\end{figure}

In order to compare this prediction with experiment, we have
estimated the exchange-enhanced effective g-factors $g_{ex}^{*}$.
Starting from the density of states, $D(E)$, as described in
Section \ref{SectionIII}, we use a simple zero temperature model,
in which $D(E)$ is described by a set of Gaussian Landau levels
with magnetic field-independent width.\cite{Piot} The longitudinal
resistivity $\rho_{xx}$ is calculated using $\rho_{xx} \propto
D(E_{F})$. We have extracted the value of $g_{ex}^{*}$ by fitting
odd filling factor minima in $\rho_{xx}$ assuming that the Landau
level broadening involved here is the same the one extracted from
the low field Shubnikov de Haas oscillations. This assumption,
also used in Ref.[\onlinecite{Leadley98}], supposes that the field
independence of the Landau level width, observed at low field,
still holds in the magnetic field region where spin-splitting
first appears. This field independent Landau level width is
suggested by the $1/B^{2}$ dependence of $\ln(\Delta \rho /
\rho_{0})$ in the Dingle plots (Fig.\ref{fig3}).

As expected, a single electron picture, in which the spin gap is
taken to be equal to the Zeeman gap, fails to reproduce the
observed spin splitting. It is however, possible to reproduce odd
filling factor minima using a magnetic field (filling factor)
dependent phenomenological spin gap, $g_{ex}^{*}\mu_{B}B$, where
$g_{ex}^{*}$ takes into account the exchange enhancement of the
spin gap. Representative results can be seen Fig.\ref{fig7}, where
we plot $\rho_{xx}(B)$, measured at $T=50$~mK, for one of the
samples. The broken lines are the simulated $\rho_{xx}(B)$
calculated for each odd filling factor using the $g_{ex}^{*}$
indicated in the legend.  The exchange-enhanced effective g-factor
collapses when approaching the non-split $\rho_{xx}$ region,
consistent with the Fogler and Shklovskii phase
transition,\cite{Fogler95,Leadley98} reflecting the
disorder-induced destruction of spin polarization at odd filling
factor.

\begin{figure}[tbp]
\includegraphics[width=0.9\linewidth,angle=0,clip]{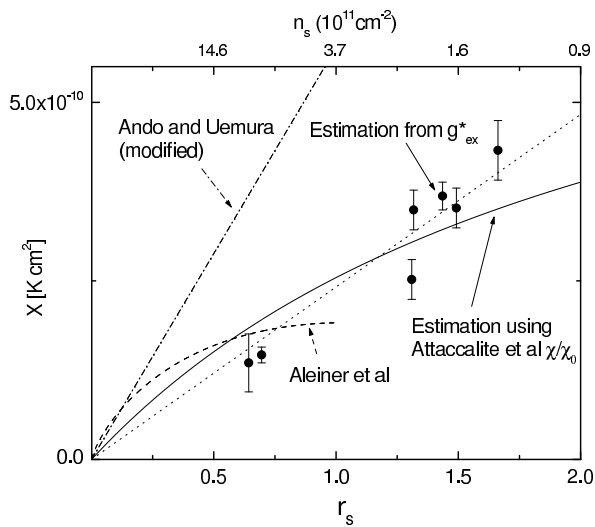}
\caption{\label{fig8} Exchange parameters as a function of the
density parameter $r_{s}$, and electron density $n_{s}$ (note
nonlinear scale for the top axis ). Estimation from Ando and
Uemura\cite{Ando74} (dash-dotted line), from Aleiner \textit{et
al.}\cite{Aleiner95}(dashed line) and from the experimentally
determined $g_{ex}^{*}$ (full circles). The dotted line is a
linear fit to $g^{*}_{ex}(r_{s})$.}
\end{figure}

When the spin gap is fully open, the spin polarization is a
maximum and the value of $g_{ex}^{*}$ (reported in table
\ref{table1}) can be compared with theory. Neglecting the single
particle Zeeman energy, we can equate the exchange gap, $X(eB/h)$,
to the total spin gap $g_{ex}^{*}\mu_{B}B$. This corresponds to
the fully polarized case in which a spin interacts with the $eB/h$
electrons in the lower spin Landau level. The values of
$X=g_{ex}^{*}\mu_{B}(h/e)$ for the different samples studied are
plotted in Fig.\ref{fig8} (closed circles) together with the
predicted value $X_{\infty}$ obtained by Eq.(\ref{eq11}) (solid
line). As previously observed in Al$_{x}$Ga$_{1-x}$As/GaAs (see
e.g. Ref.[\onlinecite{Leadley98}]), $g_{ex}^{*}$ is larger in
low-density samples. The estimation derived by Eq.(\ref{eq11})
from Attaccalite \textit{et al.} spin susceptibility is in good
quantitative agreement with the one obtained from our
$g^{*}_{ex}$, even though the exact functional dependence upon
$r_{s}$ (sub-linear) is not well reproduced by the data.

For comparison, we also show in Fig.\ref{fig8} other theoretical
estimations of the exchange parameter. Using the Ando and
Uemura\cite{Ando74} exchange-enhancement of the spin gap, we can
again write $X(eB/h)=\frac{1}{\nu}(e^{2}k_{F}/4\pi\epsilon)$. Here
the magnetic length has been replaced by $1/k_{F}$ which is the
relevant length scale at high filling
factors.\cite{Aleiner95,Leadley98}. The filling factor
$\nu=hn_{s}/eB$ and $k_{F}=\sqrt{2\pi n_{s}}$ so that in
Fig.\ref{fig8} we plot $X=e^{2}\sqrt{2\pi/n_{s}}/4\pi\epsilon$
(dotted-dashed line). The dependence on $1/k_{F}$ rather than
$l_{B}$ leads to a $1/\sqrt{n_{s}}\propto r_{s}$ density
dependance. It is striking how this simple argument describes
qualitatively the density dependance of our enhanced g-factor
$g^{*}_{ex}\propto r_{s}$ (as visible on the linear fit to
$g^{*}_{ex}(r_{s})$ in dotted line). We have also estimated the
exchange parameter from the $\alpha$ parameter describing the
exchange gap in Ref.[\onlinecite{Aleiner95}], which is valid for
$r_{s}<1$ (dashed line). The rigorous calculation,\cite{Aleiner95}
including screening, introduces a logarithmic correction which
reduces the exchange parameter at higher $r_{s}$, improving
agreement with experiment.

Finally we note that the influence of disorder on the measured
$g^{*}_{ex}$ seems to be weak since different samples with
different Landau level broadening keep within the density trend.
This is consistent with the fact that the effect of disorder on
the spin susceptibility extracted by the tilted field method is
known to be weak.\cite{Pudalov} To summarize, the good agreement
observed between experiment and theory, means that the exchange
parameter for our samples can be reliably estimated using
Attaccalite \textit{et al.} spin susceptibility. To a good
approximation, the $X_{N}$ appearing in Eq.(\ref{eq4}) can be
identified with its zero field value, $X_{\infty}$, given by
Eq.(\ref{eq11}). The absence of a magnetic field dependence can be
understood, since at high filling factors (low B) the average
electronic separation is $1/k_{F}$, rather than the magnetic
length.

\section {CRITICAL MAGNETIC FIELD FOR SPIN SPLITTING}\label{SectionV}

\subsection {Predicted field for spin splitting}

From Eq.(\ref{eq4}) in section \ref{SectionII}, we can directly
obtain a relation which explicitly gives the magnetic field
$B_{ss}$ at which the spin degeneracy of a Landau level should be
lifted, as a function of the Landau level broadening $\Gamma$, and
the electron density $n_{s}$,
\begin{equation}\label{eq12}
\\ B_{ss}
= \frac{h\Gamma\sqrt{\pi}}{e} \frac{1}{X_{\infty}(n_{s})},
\end{equation}
where $X_{\infty}$ is given by Eq.(\ref{eq11}). We see here that
$B_{ss}$ is predicted to increase with $\Gamma$, so that spin
splitting appears later in more disordered samples, which is also
experimentally well-established. The same $\Gamma$-dependance is
found for the condition for the collapse of spin splitting when
comparing the spin gap to the Landau level
broadening.\cite{Fogler95,Leadley98} The electron density
dependence of $B_{ss}$ arises via the dependence of $X_{\infty}$
on $n_{s}$. With $X_{\infty}$ defined as in Eq.(\ref{eq11}) this
gives for $B_{ss}$ a dependence close to $\sqrt{n_{s}}$, pushing
spin splitting to higher magnetic field for higher density
samples.

\subsection {Comparison with experiment}

Experimentally, the spin splitting phenomenon is probed by the
presence of a minimum appearing at odd filling factor in the
longitudinal resistance. A method of extracting $B_{ss}$ from the
experimental data is needed for a quantitative comparison. In
Ref.[\onlinecite{Fogler95}], the critical filling factor (magnetic
field) for collapse of spin splitting is defined as the filling
factor corresponding to $\delta\nu=0.5$. Here $\delta\nu$, is the
filling factor difference between two consecutive $\rho_{xx}$
maxima related to spin up and down Landau levels. This method, has
been successfully applied to analyze the ``collapse'' of
spin-splitting, in which the spin gap is compared to the disorder
in order to determine when the spin splitting should
collapse.\cite{Leadley98}

In a similar manner, we have analyzed our $\rho_{xx}(B)$ data to
extract $\delta\nu$, as shown in Fig.\ref{fig2} for sample
NU1783a. In our approach, in which we start from the
``paramagnetic'' state, the condition $\delta\nu=0$ is more
appropriate. As the condition $\delta\nu=0$ cannot be accessed
experimentally, we have to extrapolate from the experimental
$\delta\nu(B)$, for which we typically have data in the regime
$0.5\leq\delta\nu\leq1$ (see Fig.\ref{fig2}). The value of
$B_{ss}$ determined in this manner is only slightly smaller than
if we had used the $\delta\nu=0.5$ criterium, and, the error
implied by such a treatment is estimated for each sample from the
width of the transition.

Rigourously, as our model is developed for the zero temperature
case, what is required is the extrapolated value of $B_{ss}$ at
zero temperature. However, we have performed this analysis at
different temperatures ($0.05 \leq T \leq 1.2$~K)  and the
extracted evolution of $B_{ss}$ versus temperature shows that the
difference between the extrapolated $T=0$~K value, and the 50~mK
value, never exceeds 5\%, which is well within the error bar for
$B_{ss}$.

\begin{figure}[tbp]
\includegraphics[width=0.9\linewidth,angle=0,clip]{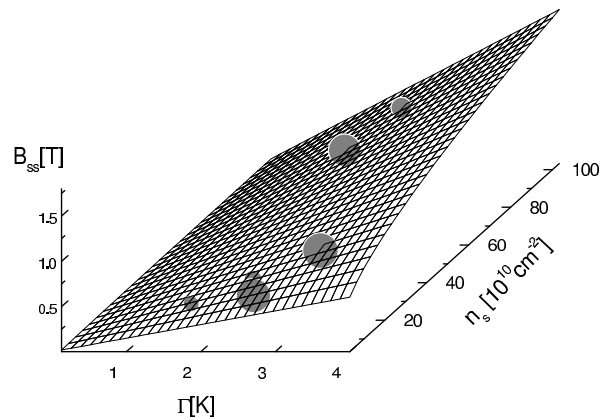}
\caption{\label{fig9} Magnetic field for spin splitting ($B_{ss}$)
plotted versus Landau level broadening ($\Gamma$), and, density
($n_{s}$). The wire frame is the prediction of Eq.(\ref{eq12}),
and the spheres are the experimentally determined $B_{ss}$. The
experimental error is indicated by the size of the spheres.}
\end{figure}

In Fig.\ref{fig9}, $B_{ss}$ is plotted both versus Landau level
broadening ($\Gamma$) and electron density ($n_{s}$),  for the six
different samples (HJ or QW), spanning a density and mobility
range respectively of $\sim (1.5-9) \times 10^{11}$~cm$^{-2}$ and
$\sim (0.1-6) \times 10^{6}$~cm$^{2}/$Vs. The $\Gamma$ values,
which are also given in Table \ref{table1}, have been estimated as
explained in Section \ref{SectionIII}. To compare experiment with
our model, the predicted evolution of $B_{ss}$ calculated using
Eq.(\ref{eq12}), is plotted as a wire mesh (3D plot) in
Fig.\ref{fig9}. The exchange energy $X_{\infty}$ has been
estimated using the calculated spin susceptibility of Attaccalite
\textit{et al.}\cite{Attaccalite} as explained in Section
\ref{SectionIV}. Fig.\ref{fig9} shows a good quantitative
agreement between the prediction and the experimental $B_{ss}$,
especially considering that there are no adjustable parameters in
the model . If we focus on the $B_{ss}$ variation at constant
disorder, the density dependence is correctly described. Equally,
at fixed density, the disorder dependance is roughly linear in
$\Gamma$ as predicted by Eq.(\ref{eq12}).

The good quantitative agreement gives further weight to the main
idea of our model in which there is a competition between the
energy cost of a flipping spins, which increases with $\Gamma$,
and the exchange gain of flipping spins, which increases with
decreasing $n_{s}$, because of stronger exchange interactions at
lower densities. The assumption that spin splitting in GaAs is
driven primarily by exchange interactions, the Zeeman splitting
being only a correction, is also validated.

\begin{figure}[t]
\includegraphics[width=0.75\linewidth,angle=0,clip]{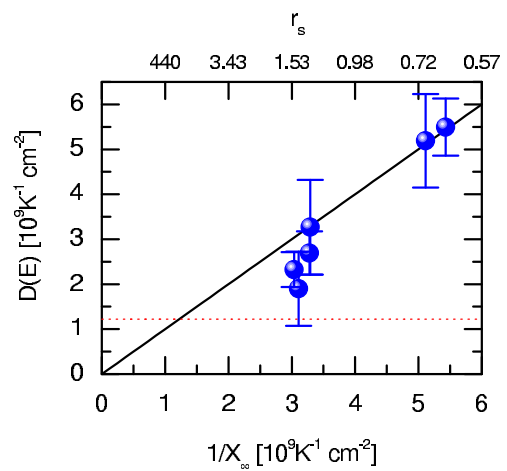}
\caption{\label{fig10} Critical density of states at the Fermi
level ($D_{ss}(E_{F})$) for a paramagnetic-ferromagnetic
transition plotted as a function of the inverse exchange energy
$1/X_{\infty}$. The $D_{ss}(E_{F})$ predicted as explained in the
text using Eq.(\ref{eq12}) (solid line) and the zero field 2D
density of states (broken line) are also plotted for comparison.}
\end{figure}

From the experimentally determined critical magnetic field,
$B_{ss}$, and the Landau level broadening, $\Gamma$, it is
possible to calculate the critical density of states at the Fermi
level, $D_{ss}(E_{F})=(eB_{ss}/h)/\sqrt{\pi}\Gamma$, necessary to
observe the paramagnetic to ferromagnetic phase transition. This
is plotted in Fig.\ref{fig10} versus the inverse exchange energy
($1/X_{\infty}$). Here, $1/X_{\infty}$ is calculated from the
electron density using Eq.(\ref{eq11}). The corresponding $r_{s}$
values are also indicated on the top axis. The predicted
dependence of $D_{ss}(E_{F})$ on $1/X_{\infty}$, calculated using
Eq.(\ref{eq12}) is plotted in Fig.\ref{fig10} for comparison. The
agreement with theory is reasonably good, although there is some
deviation at low electron density (high $r_{s}$). In these
samples, spin splitting occurs too early, before the critical
density of states is reached. This may be a correction due to the
non zero Zeeman energy. Clearly, in the limit of infinitely narrow
Landau levels, our model breaks down, and the Zeeman energy is
sufficient for the system to form a polarized ground state in the
highest Landau level, even in the absence of exchange
interactions. The density of states for a 2D system in zero
magnetic field is indicted by the dotted line in
Fig.{\ref{fig10}}. The transition to a ferromagnetic state at zero
magnetic field is predicted to occur for $r_{s}\sim26$,
corresponding to the divergence of $(\chi/\chi_{0})$, in the
calculations of Attaccalite \textit{et al.}\cite{Attaccalite}
However, the corresponding electronic density of
$n_{s}\sim5\times10^{8}$~cm$^{-2}$ is well below the density for
the metal-insulator transition in GaAs.\cite{Zhu03} At such
densities, the system can no longer be described as a paramagnetic
Fermi liquid, and hence no Stoner transition is expected at $B=0$
in GaAs, in agreement with experiment.

\section {CONCLUSION}

In conclusion, we have performed a quantitative analysis of spin
splitting in a series of GaAs heterojunctions and quantum wells. A
simple model is developed which predict that spin splitting should
occur when the density of states at the Fermi level is large
enough to stabilize a spin polarized ground state for the highest
occupied Landau level. The density of states in magnetic field has
been experimentally determined using low temperature temperature
Shubnikov de Haas measurements. The exchange strength has been
estimated from the theoretical spin susceptibility of
Ref.[\onlinecite{Attaccalite}]. Tilted-field measurements have
been used to show that the measured spin susceptibility for our
samples is in good agreement with this theory. In addition, the
calculated exchange energy has been shown to be reasonable from an
experimental estimation of the exchange-enhanced g-factors. The
predicted field for the appearance of spin splitting, $B_{ss}$,
calculated in the limit of zero Zeeman energy and zero
temperature, is in good quantitative agreement with the
experimental data at mK temperature.

As the Zeeman energy plays no role in this model, the appearance
of spin splitting is simply the manifestation of an itinerant
quantum Hall ferromagnet in the highest occupied Landau level.
This can also be thought of as a Stoner transition, since, the
only role of the magnetic field is to modify the density of states
at the Fermi energy. The critical density of states,
$D_{ss}=1/{X_{\infty}}$, for the Stoner transition, is larger than
the two dimensional density of states, $m^{*}/\pi\hbar^{2}$,
consistent with the absence of a Stoner transition at zero
magnetic field.

\begin{acknowledgments}
We would like to thank A.H. MacDonald for stimulating discussions.
\end{acknowledgments}

\bibliography{PRB}

\begin{thebibliography}{32}
\bibitem[{Gau()}]{Gaussian}
\bibinfo{note}{A linear behavior of $\ln(\Delta\rho/\rho)$ versus $1/B$ can
  also result from Gaussian Landau levels with width $\propto \sqrt{B}$. This
  case, discussed in Ref.[\onlinecite{Coleridge96}], has not been developped
  here since the relevant trend of our Dingle plots is $1/B^{2}$ rather than
  $1/B$ (See Section \ref{SectionIIIA2}).}
\end{thebibliography}

\end{document}